\documentclass{ws-rv9x6}
\usepackage{subfigure}     % required only when side-by-side / subfigures are used
\usepackage{ws-rv-van}     % numbered citation/references (default)

\makeindex
\begin{document}

\chapter{Exact solutions for pairing interactions} %\label{ra_ch1}

\author {J. Dukelsky$^1$ and S. Pittel$^2$}

\address{$^1$Instituto de Estructura de la Materia. CSIC. \\
Serrano 123, 28006 Madrid, Spain. \\
$^2$Department of Physics and Astronomy and Bartol Research Institute, University of Delaware, Newark, DE 19716 USA.
}

\begin{abstract}
The exact solution of the BCS pairing Hamiltonian was found by
Richardson in 1963. While little attention was paid to this exactly
solvable model in the remainder of the 20th century, there was a
burst of work at the beginning of this century focusing on its
applications in different areas of quantum physics. We review the
history of this exact solution and discuss recent developments
related to the Richardson-Gaudin class of integrable models,
focussing on the role of these various models in nuclear physics.
\end{abstract}

\body

\section{Cooper pairs, BCS and the Richardson exact solution}\label{ra_sec1}

The first breakthrough towards a microscopic description of the
superconducting phenomenon was due to Cooper{\cite{Cooper}}, who in
1956 showed that a single pair of electrons on top of an inert Fermi
sea could be bound by an infinitesimal attractive interaction. The
search for a many-body wave function describing a fraction of
correlated and overlapping pairs mixed with a Fermi sea was a key
goal for the rest of that year. Schrieffer came up with a solution
at the beginning of 1957 and the BCS team (Bardeen, Cooper and
Schrieffer) started an intensive and fruitful collaboration to
explain quantitatively many superconducting properties from the
associated BCS wave function. This led to the famous BCS paper
\cite{BCS} which provided a complete microscopic explanation of
superconductivity.

The success of the BCS theory quickly spread to other quantum
many-body systems, including the atomic nucleus. In the summer of
1957, David Pines visited the Niels Bohr Institute and gave a series
of seminars about the yet unpublished BCS theory. Soon thereafter
Bohr, Mottelson and Pines published a paper \cite{BMP} suggesting
that the gaps observed in even-even nuclei could be due to
superconducting correlations. They noted, however, that these
effects should be strongly influenced by the finite size of the
nucleus. Since then, and up to the present, number projection and in
general symmetry restoration in the BCS and Hartree-Fock-Bogoliubov
approximations have been important issues in nuclear structure.

At the beginning of the sixties, while several groups were
developing numerical techniques for number-projected BCS
calculations {\cite{Ker, Mang}}, Richardson provided an exact
solution for the reduced BCS Hamiltonian {\cite{Rich1, Rich2}}. In
spite of the importance of his exact solution, this work did not
have much impact in nuclear physics with just a few exceptions.
Later on, his exact solution was rediscovered in the framework of
ultrasmall superconducting grains \cite{Grain} where BCS and
number-projected BCS were unable to describe appropriately the
crossover from superconductivity to a normal metal as a function of
the grain size. Since then, there has been a flurry of work
extending the Richardson exact solution to families of
exactly-solvable models, now called the Richardson-Gaudin (RG)
models {\cite{RMP}}, and applying these models to different areas of
quantum many-body physics including mesoscopic systems, condensed
matter, quantum optics, cold atomic gases, quantum dots and nuclear
structure {\cite{Ortiz}}. In this paper, we review Richardson's
solution, its generalization to the exactly-solvable RG models and
discuss the applications of these models in nuclear physics.

\section{The Richardson solution of the reduced BCS hamiltonian}

We will focus on a pairing Hamiltonian with constant
strength $G$ acting in a space of doubly-degenerate time-reversed
states $(k,\bar{k})$,
\begin{equation}
H_P=\sum_{k } \epsilon_k c^{\dagger}_{k } c_{k } -G \sum_{k,k'}
 c^{\dagger}_{k} c^{\dagger}_{\overline{k} }
c_{\overline{k}'} c_{k'} ~,\label{HBCS}
\end{equation}
where $\epsilon _k$ are the single-particle energies for the doubly-degenerate orbits $k,\bar{k}$.

Cooper considered the addition of a pair of fermions with an
attractive pairing interaction on top of an inert Fermi sea (FS) under the influence of this Hamiltonian.
He showed that the pair eigenstate is

\begin{equation}
\left\vert \Psi _{Cooper} \right\rangle = \sum_{k>k_{F}}\frac{1}{%
2\epsilon _{k}-E} ~c_{k}^{\dagger }c_{\overline{k}}^{\dagger
}~\left\vert FS\right\rangle ~,\label{Cooper}
\end{equation}
where $E$ is the energy eigenvalue. Cooper found that for any
attractive value of $G$, the Fermi sea is unstable against the
formation of such bound pairs. Therefore, an approach that takes
into account a fraction of these correlated pairs mixed with a Fermi
sea should be able to describe the superconducting phenomenon.

The BCS approach followed a somewhat different path to the one
suggested by Cooper, defining instead a variational wave function as
a coherent state of pairs that are averaged over the whole system,

\begin{equation}
\left\vert \Psi _{BCS}\right\rangle = e^{\Gamma ^{\dagger }}\left\vert 0\right\rangle ~, \label{BCS}
\end{equation}
where $\Gamma ^{\dagger }=\sum_k z_k c^{\dagger}_{k}
c^{\dagger}_{\overline{k}}$ is the coherent pair.  Though errors due
to the non-conservation of particle number in (\ref{BCS}) are
negligible when the number of pairs is sufficiently large,  they can
be important in such finite systems as atomic nuclei {\cite{BMP}}.
To accommodate these effects, number-projected BCS (PBCS) \cite{Ker}
considers a condensate of pairs of the form

\begin{equation}
\left\vert \Psi _{PBCS}\right\rangle = \left( \Gamma ^{\dagger
}\right) ^{M}\left\vert 0\right\rangle ~,\label{PBCS}
\end{equation}
where $M$ is the number of pairs and  $\Gamma ^{\dagger }$ has the same form as in BCS.

Richardson \cite{Rich1} proposed an ansatz for the exact solution of
the pairing Hamiltonian (\ref{HBCS}) that followed closely Cooper's
original idea. For a system with $2M+\nu$ particles,  with $\nu$ of
these particles unpaired, his ansatz involves a  state of the form
\begin{equation}
\left| \Psi \right\rangle =B_{1}^{\dagger }B_{2}^{\dagger }\cdots
B_{M}^{\dagger }\left| \nu \right\rangle  ~, \label{ansa}
\end{equation}
where the collective pair operators $B^{\dagger}_{\alpha }$ have the
form found by Cooper for the one-pair problem,
\begin{equation}
B_{\alpha }^{\dagger }=\sum_{k=1}^L\frac{1}{2\varepsilon _{k}-E_{\alpha }}%
~ c_{k}^{\dagger }c_{\overline{k}}^{\dagger } ~. \label{B}
\end{equation}
Here $L$ is the number of single-particle levels and
\begin{equation}
\left\vert \nu\right\rangle \equiv \left\vert
\nu_{1},\nu_{2}\cdots,\nu_{L}\right\rangle
\end{equation}
is a state of $\nu$ unpaired fermions ($\nu=\sum_{k}\nu_{k}$, with
$\nu_{k}=1~or~0$) defined by $c_{k}c_{\overline{k}}\left\vert
\nu\right\rangle =0$, and $n_{k} \left\vert \nu\right\rangle =\nu_{
k}\left\vert \nu\right\rangle$.

In the one-pair problem, the quantities $E_{\alpha }$ that enter
(\ref{B}) are the eigenvalues of the pairing Hamiltonian, {\em
i.e.}, the {\em pair energies}. Richardson proposed to use the $M$
pair energies $E_{\alpha}$ in the many-body wave function of Eqs.
(\ref{ansa},~\ref{B}) as parameters which are chosen to fulfill the
eigenvalue equation $H_{P}\left| \Psi \right\rangle =E\left| \Psi
\right\rangle $. He showed that this is the case if the pair
energies satisfy a set of $M$ non-linear coupled equations
\begin{equation}
1-G\sum_{k=1}^{L} \frac{1-\nu_k}{2\varepsilon_{k}-E_{\alpha}}- 2G\sum_{\beta
\left( \neq \alpha \right) =1}^{M}\frac {1}{ E_{\beta}-E_{\alpha}}=0
~, \label{Richardson}
\end{equation}
which are now called the Richardson equations. The second term represents the
interaction between particles in a given pair and the third term
represents the interaction between pairs. The associated eigenvalues
of $H$ are given by
\begin{equation}
{\cal E}=\sum_{k=1}^{L} \varepsilon_{k}
\nu_{k}+\sum_{\alpha=1}^{M}E_{\alpha}~,
\end{equation}
namely as a sum of the pair energies.

Each independent solution of the set of Richardson equations defines
a set of $M$ pair energies that completely characterizes a
particular eigenstate (\ref{ansa},~\ref{B}). The complete set of
eigenstates of the pairing Hamiltonian can be obtained in this way.
The ground state solution is the energetically lowest solution in
the $\nu=0$ or $\nu=1$ sector, depending on whether the system has
an even or an odd number of particles, respectively.

There are a couple of points that should be noted here. First, in
contrast to the BCS solution, each Cooper pair
$B^{\dagger}_{\alpha}$ is distinct. Second, if one of
the pair energies $E_\alpha$ is complex, then
its complex-conjugate $E^*_\alpha$ is also a solution. From this
latter point we see that $|{\Psi}\rangle$ preserves time-reversal
invariance.

On inspection of the Richardson pair (\ref{B}), we see that a pair
energy that is close to a particular $2 \epsilon _k$, \emph{i.e.}
close to the energy of an unperturbed pair, is dominated by this
particular configuration and thus defines an uncorrelated pair. In
contrast, a pair energy that lies sufficiently far away in the
complex plane produces a correlated Cooper pair. This is to be
contrasted with the single BCS coherent pair, which has amplitude
$z_k=v_k/u_k$ and which mixes correlated and uncorrelated pairs over
the {\it whole} system.

\section{Generalization to the Richardson-Gaudin class of integrable
models}

In this section, we discuss how to generalize the standard pairing
model, which as we have seen is exactly solvable, to a wider variety
of exactly-solvable models, the so-called Richardson-Gaudin
models{\cite{Class}}, all of which are based on the $SU(2)$ algebra.
We first introduce the generators of $SU(2)$, using a basis more
familiar to nuclear structure,

\begin{equation}
K^0_j = \frac{1}{2}\left( \sum_m a^{\dagger}_{jm}a_{jm}-\Omega_j \right) ,~
K^{+}_j = \sum_m a^{\dagger}_{jm} a^{\dagger}_{j\overline{m}}
,~K^-_j=(K^+_j)^{\dagger}. \label{NA}
\end{equation}
Here $a^{\dagger}_{jm}$ creates  a fermion in single-particle state
$jm$,  ${j\overline{m}}$ denotes the time reverse of $jm$, and
$\Omega_j=j+\frac{1}{2}$ is the pair degeneracy of orbit $j$. These
operators fulfill the $SU(2)$ algebra $[K^+_j , K^-_{j'}]=2
\delta_{jj'} K^0_j $, $[K^0_j , K^{\pm}_{j'}]=\pm \delta_{jj'}
K^{\pm}_{j}$ .

We now consider a general set of $L$ Hermitian and number-conserving
operators that can be built up from the generators of $SU(2)$ with
linear and quadratic terms,

\begin{equation}
R_{i}=K_{i}^{0}+2g\sum_{j\left( \neq i\right) } \left[
\frac{X_{ij}}{2}  \left(
K_{i}^{+}K_{j}^{-}+K_{i}^{-}K_{j}^{+}\right) +
 Y_{ij}K_{i}^{0}K_{j}^{0} \right]~. \label{Rgen}
\end{equation}

Following Gaudin{\cite{Gaudin}}, we then look for the conditions
that the matrices $X$ and $Y$  must satisfy in order that the $R$
operators commute with one another. It turns out that there are
essentially two families of solutions, referred to as the rational
and hyperbolic families, respectively.
\begin{description}
\item {{\it i}. The rational family}
\begin{equation}
X_{ij}=Y_{ij}=\frac{1}{\eta _{i}-\eta _{j}} \label{R}
\end{equation}

\item {{\it ii.} The hyberbolic family}

\begin{equation}
X_{ij}=2\frac{\sqrt{\eta_i \eta_j}}{\eta_i-
\eta_j}~,~Y_{ij}=\frac{\eta _{i}+\eta _{j}}{\eta _{i}-\eta
_{j}}\label{H}
\end{equation}
\end{description}
Here the set of $L$ parameters $\eta_i$ are free real numbers.

The traditional pairing model is an example of the rational family.
It can be obtained as a linear combination of the integrals of
motion, $H_P=\sum_j \varepsilon_j R_j (\varepsilon_j)$, with $\eta_j
= \varepsilon_j$.

The complete set of eigenstates of the rational integrals of motion
is given by the Richardson ansatz (\ref{ansa},~\ref{B}). This fact
led Gaudin \cite{Gaudin} to try to relate his integrable models to
the BCS Hamiltonian without success. The proof of integrability of
the BCS Hamiltonian was found later in ref. {\cite{CRS}}.  We will
not present the general solution of the two integrable families
here, referring the reader to refs. {\cite{Class, RMP, Ortiz}}.

The key point is that any Hamiltonian that can be expressed as a
linear combination of the $R$ operators can be treated exactly using
this method. In the following sections, we discuss nuclear
applications of the standard pairing model and of a new model based
on the hyberbolic family.

\section{Applications of the Richardson solution to pairing in nuclear physics}

Richardson himself started to explore analytically the exact
solution in nuclear structure for few pairs outside a doubly-magic
core{\cite{Rich3,Rich4}}. He also proposed a numerical method to
solve the equations for systems with equidistant
levels{\cite{Rich5}}, a model that was subsequently used as a
benchmark to test many-body approximations {\cite{Bang}}. However,
the first application of the Richardson solution to a real nuclear
system was reported by Andersson and Krumlinde {\cite{Krum}} in
1977. They studied the properties of high-spin states in $^{152}Dy$
using an oblate deformed oscillator potential and including the
effects of pairing at several different levels of approximation.
They compared the results when pairing was treated with the
traditional BCS approximation, when it was treated in PBCS
approximation (using the saddle point approximation) and when it was
treated exactly using the Richardson method.

Following that early work, there were sporadic references to the
Richardson method but no realistic studies of atomic nuclei until
just a few years ago. In 2007, Dussel {\it et al.} \cite{Dussel}
reported a systematic study of pairing correlations in the even Sm
isotopes, from $^{144}Sm$ through $^{158}Sm$, using the
self-consistent deformed Hartree Fock+BCS method. The calculations
made use of the density-dependent Skyrme force, SLy4, and treated
pairing correlations using a pairing force with constant strength
$G$ assuming axial symmetry and taking into account 11 major shells.

Using the results at self-consistency to define the HF mean field,
pairing effects within {\it that} mean field were then considered
using the alternative number-conserving PBCS approach and the exact
Richardson approach. In this way it was possible to directly compare
the three approaches to pairing with the same pairing Hamiltonian, a
primary focus of the study. It should be noted here that the Hilbert
space dimensions associated with the residual neutron pairing
Hamiltonian is of the order of $3.9 \times 10^{53}$ for $^{154}Sm$,
whereas the exact Richardson approach requires the solution of a
coupled set of $46$ non-linear equations.

In the one semi-magic nucleus $^{144}Sm$ that was studied, the
principal correlation effects arise when projection is included,
taking the system from one that is normal at the level of BCS to one
with substantial pairing correlations. Treating pairing exactly
provides a further modest increase in pairing correlations of about
$0.3$ $MeV$. In non-semi-magic nuclei, the effect on the pairing
correlation energy of the exact solution is significantly more
pronounced. While there too number projection provides a substantial
lowering of the energy, it now misses about $1$ $MeV$ of the exact
correlation energy that derives from the Richardson solution.

 \begin{figure}[htb]
\includegraphics[height=.4\textheight]{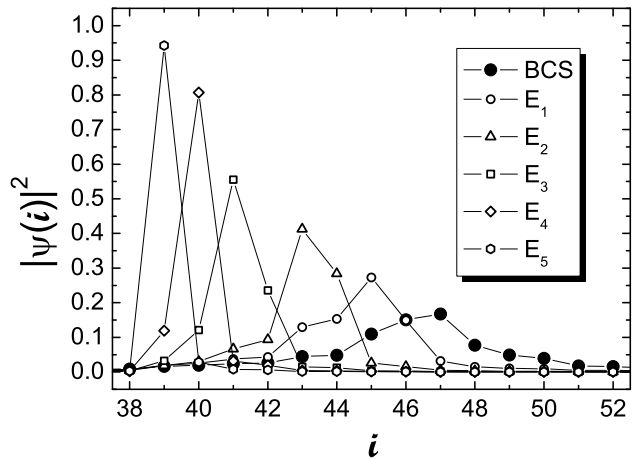}
\caption{Square of the wave function of the most collective Cooper
pairs in $^{154}$Sm (denoted E$_1$, E$_2$, E$_3$, E$_4$, and E$_5$)
and the pairing tensor (BCS) versus the single-particle level $i$ in
the Hartree-Fock basis The results are presented for the physical
value of the pairing strength, $G=0.106$ $MeV$}
\end{figure}

As noted earlier, the Richardson prescription gives rise to distinct
Cooper pairs with distinct structure. This is illustrated in Figure
1  where we compare the square of the wave function for the most
correlated Cooper pairs in $^{154}$Sm, {\it i.e.} those whose pair
energies lie farthest from the real axis in the complex plane, with
the square of the pairing tensor $u_{i} v_{i}$ that derives from the
corresponding BCS solution. All wave functions are plotted versus
the order of the single-particle states to make clear the relevant
mixing of configurations in each pair. The pair label E$_1$ refers
to the two most collective pairs (with complex conjugate pair
energies). E$_2$ refers to the next two most collective pairs, which
are however only marginally collective. E$_3$ refers to the next two
in descending order of collectivity, but they only involve
perturbative mixing of configurations and are not truly collective.
The final two that are shown, E$_4$ and E$_5$, have real pair
energies and involve almost pure single-particle configurations.
>From the figure, we see that even the most collective Cooper pairs
are much less collective than $u_{i} v_{i}$, and therefore that
their size in coordinate space is significantly larger than that of
the BCS pairing tensor{\cite{BCS-BEC}}, which is often  used in the
literature as a definition of the Cooper pair wave
function{\cite{Mat}}.

The exact Richardson solution was also used to study the gradual
emergence of superconductivity in the Sn isotopes
{\cite{Electrostatic}}. By making use of an exact mapping between
the Richardson equations and a classical electrostatic problem in
two dimensions, it was possible to get a physical picture of how
superconductivity develops as a function of the pairing strength. In
particular, as the pairing strength is increased the pair energies
 gradually merge into larger structures in the complex plane as
pair correlations gradually overcome single-particle effects.

More recently, the Richardson solution has been applied to the
treatment of pair correlations involving the continuum. The first
work by Hasegawa and Kaneko \cite{Hasegawa} considered only the
effect of resonances in the continuum and as a result obtained
complex energies even for the bound states of the system. Subsequent
work by Id Betan \cite{IdBetan1, IdBetan2} included the effects of
the true continuum. The most recent paper \cite{IdBetan2} treated
nuclear chains that include both bound and unbound systems,
\emph{e.g.} the even-A Carbon isotopes up to $^{28}C$. When the
system is bound, the pair energies that contribute to the ground
state occur in complex conjugate pairs, thus preserving the real
nature of the ground state energy. Once the system becomes unbound
this ceases to be the case. Now the pair energies that contribute to
the ground state do not occur in complex conjugate pairs, explaining
how a width arises in the energy of an unbound system within the
Richardson approach.

\section{The hyperbolic model}\label{ra_sec6}
The hyperbolic family of models did not find a physical realization
until very recently when it was shown that they could model a
$p$-wave pairing Hamiltonian in a 2-dimensional lattice
{\cite{Sierra}}, such that it was possible to study with the exact
solution an exotic phase diagram having a non-trivial topological
phase and a third-order quantum phase transition {\cite{RDO}}.
Immediately thereafter, it was shown that the hyperbolic family
gives rise to a separable pairing Hamiltonian with 2 free parameters
that can be adjusted to reproduce the properties of heavy nuclei as
described by a Gogny HFB treatment {\cite{Gogny}}. Both applications
are based on a simple linear combination of hyperbolic integrals
which give rise to the separable pairing Hamiltonian
\begin{equation}
H=\sum_{i} \eta_i K^0_i - G \sum_{i,i^\prime} \sqrt{\eta_i
\eta_{i^\prime}} K^+_i K^-_{i^\prime} .  \label{Ham}
\end{equation}

If we interpret the parameters $\eta_i$ as single-particle energies
corresponding to a nuclear mean-field potential, the pairing
interaction has the unphysical behavior of increasing in strength
with energy. In order to reverse this unwanted effect, we define
$\eta_i = 2 (\varepsilon_i - \alpha)$,
where the free parameter $\alpha$ plays the role of an energy cutoff and $%
\varepsilon_i$ is the single-particle energy of the mean-field level
$i$. Making use of the pair representation of $SU(2)$, the
exactly-solvable pairing Hamiltonian (\ref{Ham}) takes the form
\begin{equation}
H=%H^{\prime}+\alpha(2M-L_c)+2\sum_i s_i \varepsilon_i= \\
%& &
\sum_{i} \varepsilon _{i} \left( c_{i}^{\dagger }c_{i}+c^{\dagger}_{\overline{i}%
} c_{\overline{i}}\right)
 -2 G\sum_{ii^{\prime }}\sqrt{\left( \alpha -\varepsilon _{i}\right)
\left( \alpha -\varepsilon _{i^{\prime }}\right) }~c_{i}^{\dagger }c_{%
\overline{i}}^{\dagger }c_{\overline{i}^{\prime }}c_{i^{\prime }},
\label{Hint}
\end{equation}
with eigenvalues $E=2\alpha M+\sum_i \varepsilon_i \nu_i
% \langle\nu|c_{i}^{\dagger }c_{i}+c_{\overline{i}%
%}^{\dagger} c_{\overline{i}} |\nu\rangle
+ \sum_{\beta} E_{\beta} $. The pair energies $E_{\beta}$ correspond
to a solution of the set of non-linear Richardson equations

\begin{equation}
\frac{1}{2} \sum_{i}\frac{1}{\eta _{i}-E_{\beta }}-\sum_{\beta ^{\prime
}(\not=\beta )}\frac{1}{E_{\beta ^{\prime }}-E_{\beta }}=\frac{Q}{%
E_{\beta }}~,  \label{eq:sp_eigeneq}
\end{equation}
where $Q=\frac{1}{2 G}-\frac{L}{2}+M-1$. Each particular solution
of Eq. (\ref{eq:sp_eigeneq}) defines a unique eigenstate.

\begin{figure}
\centerline{\psfig{file=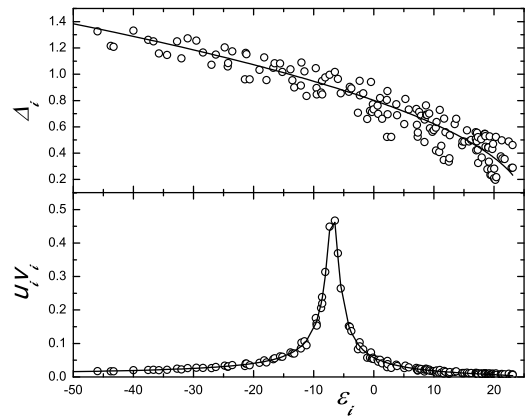,width=8cm}} \caption{Gaps
$\Delta_i$ and pairing tensor $u_i v_i$ for protons in $^{238}U$.
Open circles are Gogny HFB results in $MeV$. Solid lines are BCS
results of the hyperbolic Hamiltonian in $MeV$.} \label{U}
\end{figure}

Due to the separable character of the hyperbolic Hamiltonian, in
BCS approximation the gaps  $\Delta_i= 2 G \sqrt{ \alpha -\varepsilon_{i}} \sum_{i^{\prime}} \sqrt{%
\alpha -\varepsilon_{i^{\prime}}} u_{i^{\prime}}v_{i^{\prime}} =
\Delta \sqrt{ \alpha -\varepsilon_{i}}$ and the pairing tensor $ u_i
v_i= \frac{\Delta \sqrt{\alpha-\varepsilon_{i}}}
{2\sqrt{(\varepsilon_{i} -\mu)^2+ (\alpha -\varepsilon_{i})
\Delta^2}}$ have a very restricted form. In order to test the
validity of the exactly solvable Hamiltonian (\ref{Hint}) we take
the single-particle energies $\varepsilon_i$ from the HF energies of
a Gogny HFB calculation and we fit the parameters $\alpha$ and $G$
to the gaps and pairing tensor in the HF basis. Figure \ref{U} shows
the comparison for protons in $^{238}U$ between the Gogny HFB
results in the HF basis and the BCS approximation of the hyperbolic
model. From these results we extracted the values $\alpha=25.25$ $ MeV$ and
$G=2 \times 10^{-3}MeV$. The valence space determined by the cutoff
$\alpha$ corresponds to 148 levels with 46 proton pairs. The size of
the Hamiltonian in this space is $4.83 \times 10^{38}$, well beyond
the limits of exact diagonalization. However, the integrability of
the hyperbolic model provides an exact solution by solving a set of
46 non-linear coupled equations. Moreover, the exact solution shows
a gain in correlation of more than 2 $MeV$ suggesting the importance
of taking into account correlations beyond mean-field.

\section{Extensions to non-compact and higher rank algebras}

Up to now, we have restricted our discussion to RG models that are
based on the compact rank-1 $SU(2)$ pair algebra. The method of
constructing RG models can be extended to the non-compact rank-1
$SU(1,1)$ algebra as well, whereby pairing in bosonic
systems\cite{bosons} is described in complete analogy with the
$SU(2)$ case. An early application to the $SO(6)$ to $U(5)$ line of
integrability of the Interacting Boson Model (IBM) was reported in
ref. {\cite{Draayer}}, with the exact solution being obtained there
directly using an infinite dimensional algebraic technique. Further
work on the IBM using the integrable $SU(1,1)$ RG model\cite{PRLIBM}
including high-spin bosons ($d$, $g$, $\cdots$) revealed a
particular feature of the repulsive boson pairing interaction that
seems to provide a new mechanism for the enhancement of $s-d$
dominance, giving further support for the validity of the $s-d$
Interacting Boson Model.

The RG models are not constrained to rank-1 algebras. They can be
extended to any semi-simple Lie algebra {\cite{Aso}}. Richardson
himself studied some restricted solutions of the T=1 pairing model
\cite{Rich6} and the T=0,1 pairing model {\cite{Rich7}}. As a
general statement, the reduced pairing Hamiltonian is exactly
solvable for any multi-component system. The first step in finding
an exact solution is to identify the Lie algebra of the commuting
pair operators and then to specialize the general solution given in
{\cite{Aso}}.  One has to keep in mind that while the $SU(2)$ RG
model has a single set of unknown parameters, the pair energies,
larger rank algebras have as many sets of unknown parameters as the
rank of the algebra. Therefore, the higher the rank of the algebra,
the greater is the complexity of the solution. Several pairing
Hamiltonians with relevance to nuclear physics have been studied in
the last few years.

{\it i.} The rank-2 $SO(5)$ RG model\cite{T1} describes T=1
proton-neutron pairing with non-degenerate single particle levels.
The exact solution has two sets of spectral parameters, the pair
energies and a second set associated with the $SU(2)$ isospin
subalgebra. In spite of the greater complexity, it was possible to
solve exactly a T=1 pairing Hamiltonian for the nucleus $^{64}Ge$
using a $^{40}Ca$ core, with a Hilbert space dimension well beyond
the limits of exact diagonalization.

{\it ii.} The rank-3 $SO(6)$ RG model \cite{Color} describes color
pairing, {\it i. e.} pairing between three-component fermions. The
exact Richardson equations have three sets of spectral parameters,
of which one correspond the the pair energies and the other two are
responsible for the different couplings within the $SU(3)$ color
subalgebra. The model has been used to study the phase diagram of
polarized three-component fermion atomic gases. However, it could in
principle be exploited to describe non-relativistic quark systems.

{\it iii.} With increasing complexity, the rank-4 $SO(8)$ RG model
\cite{ST} describes either T=0,1 proton-neutron pairing or
four-component fermion gases. It contains four sets of spectral
parameters. The model has been used to study alpha-like structures
represented by clusters in the parameter space, and how these
clusters dissolve into like-particle pairs with increasing isospin.

{\it iv.} The rank-2 non-compact $SO(3,2)$ algebra generalizes the
bosonic RG models to systems of interacting proton and neutron
bosons{\cite{IBM2}}. The model describes the IBM2 in the line of
integrability between vibrational and $\gamma$-soft nuclei. The
exact solution has been employed to study the influence of high-spin
$f$ and $g$ bosons in the low-energy spectrum

\section{Summary and future outlook}
In this article, we have reviewed Richardson's solution of the
pairing model and have discussed its generalization to a wider class
of exactly solvable models. We have also discussed the application
of these models to a variety of problems in nuclear structure
physics in which pairing plays a role. It should be noted here,
however,  that all of the models that we have discussed are
restricted to the pairing degree of freedom and thus do not allow
explicit treatment of deformation effects. It is only through the
use of Nilsson or deformed Hartree Fock single-particle energies
that effects of deformation are simulated.

A key feature of the Richardson-Gaudin integrable models, is
that they transform the diagonalization of the hamiltonian matrix, whose dimension grows exponentially with the size of the system,
to the solution of a set of $M$ coupled non-linear equations where $M$ is the number of pairs.
This makes it possible to treat
problems that could otherwise not be treated and in doing so to
obtain information that is otherwise inaccessible.  For example, we
reported an application of the rational RG pairing model to the
even-mass Sm isotopes, where the size of the Hilbert space would
exceed $10^{53}$ states, and an application of the hyperbolic 
RG pairing model to $U^{238}$, where the size of the Hilbert space would
exceed $10^{38}$ states. In both cases, substantial gains in
correlation energy were found when the problem was treated exactly.

The exactly solvable RG Hamiltonians also provide excellent
benchmarks for testing approximations beyond HFB in realistic
situations both for even-even and odd-mass nuclei. Moreover, a
self-consistent HF plus exact pairing approach could in principle be
implemented to describe large regions of the table of nuclides. It
might be possible to extend such a self-consistent approach to the
O(5) RG model, providing in this way a better
description of those nuclei with $N \sim Z$ in which T=1 proton-neutron 
pairing correlations are expected to play a significant role. Unfortunately, the SO(8)
T=0,1 RG model cannot accommodate the spin-orbit splitting in the
single-particle energies.
 Nevertheless, this model could play an
important role in helping to understand quartet clusterization and
quartet condensation in nuclear and cold atom systems. Finally,
extension of the RG models to include the effects of the continuum
seems to be an especially promising avenue to explore the physics of
weakly-bound nuclei.

\section*{Acknowledgments}
This work was supported in part by the Spanish Ministry for Science
and Innovation under Project No. FIS2009-07277 and the National
Science Foundation under Grant No. PHY-0854873.


\begin{thebibliography}{9}

\bibitem{Cooper}
   L. N. Cooper, Bound Electron Pairs in a Degenerate Fermi Gas,
\emph{Phys. Rev.} {\bf 104}, 1189--1190, (1956).


\bibitem{BCS}
   J. Bardeen, L. N. Cooper, and J. R. Schrieffer, Theory of Superconductivity,
   \emph{Phys. Rev.} {\bf 108}, 1175--1204, (1957).

\bibitem{BMP}
   A. Bohr, B. R. Mottelson, and D. Pines, Possible Analogy between
the Excitation Spectra of Nuclei and those of the Superconducting
Metallic State,
   \emph{Phys. Rev.} {\bf 110}, 936--938, (1958).

\bibitem{Ker}
   A. K. Kerman, R. D. Lawson, and M. H. Macfarlane, Accuracy of the
Superconductivity Approximation for Pairing Forces in Nuclei,
   \emph{Phys. Rev.} {\bf 124}, 162--167, (1961).

\bibitem{Mang}
K. Dietrich, H. J. Mang, and J. H. Pradal, Conservation of Particle
Number in the Nuclear Pairing Model,
 \emph{Phys. Rev.} {\bf 135}, 22--34, (1964).

\bibitem{Rich1}
   R. W. Richardson, A Restricted Class of Exact Eigenstates of the
Pairing-Force Hamiltonian,
   \emph{Phys. Lett.} {\bf 3}, 277--279, (1963).

\bibitem{Rich2}
R. W. Richardson, Exact Eigenstates of Pairing-Force Hamiltonian,
   \emph{Nucl. Phys.} {\bf 52}, 221--238, (1964).

\bibitem{Grain}
 G. Sierra, J. Dukelsky, G. G. Dussel, J. von Delft, and F. Braun,
 Exact study of the effect of level statistics in ultrasmall superconducting grains,
\emph{Phys. Rev. B } {\bf 61}, 11890--11893, (2000).


\bibitem{RMP}
J. Dukelsky, S. Pittel, and G. Sierra, Exactly solvable
Richardson-Gaudin models for many-body quantum systems, \emph{Rev.
Mod. Phys.} {\bf 76}, 643--662 (2004).

\bibitem{Ortiz}
G. Ortiz, R. Somma, J. Dukelsky, and S. Rombouts, Exactly-solvable
models derived from a generalized Gaudin algebra, \emph{Nucl. Phys.
B} {\bf 707}, 421--457 (2005).


\bibitem{Class}
J. Dukelsky, C. Esebbag, and P. Schuck, Class of Exactly Solvable
Pairing Models, \emph{Phys. Rev. Lett} {\bf 87}, 066403 1--4 (2001).

\bibitem{Gaudin}
M. Gaudin, Diagonalization of a Class of Spin Hamiltonian, \emph{J.
Phys. (Paris)} {\bf 37}, 1087--1098 (1976).

\bibitem{CRS}
M. C. Cambiaggio,A. M. F. Rivas, and M. Saraceno, Integrability of
the pairing Hamiltonian,
 \emph{Nuc. Phys. A} {\bf 624}, 157--167 (1997).

\bibitem{Rich3}
R. W. Richardson, Application to the Exact Theory of the Pairing
Model to some Even Isotopes of Lead,
   \emph{Phys. Lett.} {\bf 5}, 82--84, (1964).


\bibitem{Rich4}
R. W. Richardson and N. Sherman, Pairing Models of $^{206}Pb$,
$^{204}Pb$ and $^{202}Pb$,
   \emph{Nuc. Phys.} {\bf 52}, 253--268, (1964).

\bibitem{Rich5}
R. W. Richardson, Numerical Study of 8-32 Particle Eigenstates of
Pairing Hamiltonian,
   \emph{Phys. Rev.} {\bf 141}, 949--956, (1966).

\bibitem{Bang}
J. Bang and J. Krumlinde, Model Calculations with Pairing Forces,
\emph{Nucl. Phys. A} {\bf 141}, 18--32, (1970).

\bibitem{Krum}
C. G. Andersson and J. Krumlinde, Oblate High-Spin Isomers,
\emph{Nucl. Phys. A} {\bf 291}, 21--44, (1977).

\bibitem{Dussel}
G. G. Dussel, S. Pittel, J. Dukelsky, P. Sarriguren, Cooper pairs
in atomic nuclei, \emph {Phys. Rev. C} {\bf 76} 011302 1--5, (2007).


\bibitem{BCS-BEC}
Similar results has been obtained for a 3D homogeneous diluted Fermi gas in the BCS phase.
G. Ortiz and J. Dukelsky, BCS-to-BEC crossover from the exact BCS solution,
\emph {Phys. Rev. A} {\bf 72} 043611  1--5, (2005).

\bibitem{Mat}
M. Matsuo, Spatial structure of neutron Cooper pair in low density uniform matter,
\emph {Phys. Rev. C} {\bf 73} 044309  1--16, (2005); and Matsuo's contribution to this Volume.

\bibitem{Electrostatic} J. Dukelsky, C. Esebbag, S. Pittel, Electrostatic
mapping of nuclear pairing, \emph{ Phys. Rev. Lett.} {\bf 88} 062501
1--4, (2002).

\bibitem{Hasegawa}
M. Hasgawa and K. Kaneko, Effects of resonant single-particle states
on pairing correlations, \emph{Phys. Rev. C} {\bf 67}, 024304 1--4,
(2003).

\bibitem{IdBetan1}
R. Id Betan, Using continuum level density in the pairing
Hamiltonian: BCS and exact solutions, \emph{Nucl. Phys. A} {\bf 879}
14--24, (2012).

\bibitem{IdBetan2}
R. Id Betan, Exact eigenvalues of the pairing Hamiltonian using
continuum level density, Nucl-th arxiv:1202.3986 (2012).


\bibitem{Sierra} M. Iba\~{n}ez, J. Links, G. Sierra, and S-Y Zhao,
Exactly solvable pairing model for superconductors with px+ipy-wave
symmetry, \emph{Phys. Rev. B} {\bf 79} 180501 1--4, (2009).

\bibitem{RDO} S. M. A. Rombouts, J. Dukelsky, and G. Ortiz, Quantum
phase diagram of the integrable px+ipy fermionic superfluid,
 \emph{Phys. Rev. B}  {\bf 82} 224510 1--4, (2010).


\bibitem{Gogny}  J. Dukelsky, S. Lerma H., L. M. Robledo, R.
Rodriguez-Guzman,  and S. M. A. Rombouts, Exactly solvable
Hamiltonian for heavy nuclei,
 \emph{Phys. Rev. C}  {\bf 84} 061301 1--4, (2011).

\bibitem{bosons}
J. Dukelsky and P. Schuck, Condensate Fragmentation in a New Exactly Solvable Model for Confined Bosons, \emph{Phys. Rev. Lett.}  {\bf 86}, 4207--4210 (2001).

\bibitem{Draayer}
Feng Pan and J.P. Draayer, New algebraic solutions for SO(6) to U(5) transitional nuclei in the Interacting Boson Model, \emph{Nucl. Phys. A} {\bf 636}, 156-168 (1998).

\bibitem{PRLIBM}
J. Dukelsky and S. Pittel, New Mechanism for the Enhancement of sd Dominance in Interacting Boson Models,
\emph{Phys. Rev. Lett.}  {\bf 86}, 4791--4794 (2001).


\bibitem{Aso}
M. Asorey, F. Falceto, and G. Sierra, Chern–Simons theory and BCS
superconductivity,
 \emph{Nucl. Phys. B} {\bf 622}, 593--614, (2002).

\bibitem{Rich6}
R. W. Richardson, Eigenstates of the J=0 T=1 Charge-Independent
Pairing Hamiltonian, \emph{Phys. Rev.} {\bf 144}, 874--883, (1966).

\bibitem{Rich7}
R. W. Richardson, Eigenstates of the L=0 T=1 Charge- and
Spin-Independent Pairing Hamiltonian, \emph{Phys. Rev.} {\bf 159},
792--805, (1967).

\bibitem{T1}
J. Dukelsky, V. G. Gueorguiev, P. Van Isacker, S. Dimitrova, B.
Errea, and S. Lerma H., Exact Solution of the Isovector
Neutron-Proton Pairing Hamiltonian, \emph{Phys. Rev. Lett.} {\bf
76}, 072503 1--4, (2006).


\bibitem{Color}
B. Errea, J. Dukelsky, and G. Ortiz, Breached pairing
in trapped three-color atomic Fermi gases, \emph{Phys. Rev. A.} {\bf
79}, 051603 1--4, (2009).

\bibitem{ST}
S. Lerma H., B. Errea, J. Dukelsky, and W. Satula, Exact Solution of
the Spin-Isospin Proton-Neutron Pairing Hamiltonian, \emph{Phys.
Rev. Lett.} {\bf 79}, 032501 1--4 (2007).

\bibitem{IBM2}
S. Lerma H., B. Errea, J. Dukelsky, S. Pittel,and P. Van Isacker,
Exactly solvable models of proton and neutron interacting bosons,
\emph{Phys. Rev. C}  {\bf 74} 024314 1--7, (2011).


\end{thebibliography}
\end{document}